\documentstyle[12pt,aaspp4]{article}

\def\gtsim{\ {\raise-0.5ex\hbox{$\buildrel>\over\sim$}}\ }
\def\ltsim{\ {\raise-0.5ex\hbox{$\buildrel<\over\sim$}}\ }
\slugcomment{Submitted to The Astronomical Journal}

\begin{document}

\title{
The Stellar Populations of NGC~3109:\\
Another Dwarf Irregular Galaxy with a Population II Stellar Halo}

\author{Dante Minniti $^{1,2}$, Albert A. Zijlstra $^{3,4}$ \& M. Victoria Alonso$^{5}$}

\altaffiltext{1}{Lawrence Livermore National Laboratory, MS L-413, P.O. Box 808,Livermore, CA 94550\\
E-mail:  dminniti@llnl.gov}

\altaffiltext{2}{Departamento de Astronomia, P. Universidad Cat\'olica, Casilla 104, Santiago 22, Chile}

\altaffiltext{3}{Department of Physics, UMIST, P.O.Box 88, Manchester M60 1QD, UK\\
E-mail: Albert.Zijlstra@umist.ac.uk}

\altaffiltext{4}{European Southern Observatory, Karl-Schwarzschild-Str. 2, D-85748 Garching b.  M\"{u}nchen, Germany}

\altaffiltext{5}{Observatorio Astron\'omico de C\'ordoba, Laprida 854, 5000 C\'ordoba, Argentina\\
E-mail: vicky@oac.uncor.edu}

\begin{abstract}
We have obtained V and I-band photometry for about 17500 stars
in the field of the dwarf irregular galaxy NGC~3109, located in the
outskirts of the Local Group. The photometry allows us to study the stellar
populations present inside and outside the disk of this galaxy.
{}From the VI color-magnitude diagram we
infer metallicities and ages for the stellar populations in the main
body and in the halo of NGC~3109. The stars in the disk of this
galaxy have a wide variety of ages, including very young stars with
$\sim 10^{7}$ yr. Our main result is to
establish the presence of a halo consisting of
population II stars, extending out to about 4.5 arcmin (or 1.8 kpc)
above and below the plane of this galaxy. 
For these old stars we derive an age of $>
10^{10}$ yr and a metallicity of $[Fe/H] = -1.8 \pm 0.2$. We
construct a deep luminosity function, obtaining an accurate distance
modulus $(m-M)_0 = 25.62 \pm 0.1$ for this galaxy based on the I-magnitude
of the red giant branch (RGB) tip and adopting $E(V-I) = 0.05$.  
\end{abstract}

\keywords{Galaxies: individual (NGC~3109, DDO~236) -- 
Galaxies: stellar content -- Galaxies: irregular -- 
Local Group -- Galaxy: formation}

\section{Introduction}
The results of the MACHO Project indicate that about $50^{+30}_{-20}$\% of
the dark matter in the Milky Way halo is  baryonic, and made
of faint objects with typical masses around $0.5^{+0.3}_{-0.2} ~M_{\odot}$
(Alcock et al. 1997). One possibility is that these objects are
stellar remnants from old populations, even though their initial
mass function may be extreme (Chabrier et al. 1996). If some stars 
form before galaxies, they would remain collisionless, populating
the galactic halos just like elementary cold dark matter particles 
(Miralda-Escud\'e \& Rees 1997, Loeb 1997).  If this is indeed the case, 
a simple question to answer would be: Are there old populations of 
stars observed in the halos of other galaxies which, on dynamical grounds, 
are known to have large amounts of dark matter? In order to answer 
this question, some obvious galaxies to target are the dwarf irregulars. 
These galaxies are known to be dark matter dominated, having
rising rotation curves that extend beyond their optical
limits (Carignan \& Freeman 1988). A typical case is NGC~3109,
which is located in the outskirts of the Local Group 
(van den Bergh 1994, Mateo 1998). This galaxy
is far enough not to have been affected by interactions with the 
larger Local Group spirals, yet
close enough that it can be resolved into stars. 

NGC 3109 has been relatively well studied: some of its parameters are listed
in Table 1 (see Mateo 1998 and references therein).
We note that NGC~3109 is on the bright end of the family of
Local Group dwarf irregulars, almost as luminous as the LMC (Table 1).
This galaxy has a system of globular clusters (Demers et al. 1985).
Lee (1993) showed the differences between the color-magnitude diagrams
of the NGC~3109 disk and a field located 2 arcmin away from it. 
The outer field seems to
be dominated by an old and metal-poor population. 
Earlier, Sandage (1971) pointed out that direct photographs of
all galaxies in the Local Group reveal the presence of a background sheet 
of red stars, and that the brightest of these stars are at the tip
of the red giant branch of a globular cluster-like population.
Based on these pieces of evidence, and motivated by the
discovery of a halo made of Population II stars in the dwarf irregular
galaxy WLM (Minniti \& Zijlstra 1996), we decided  to investigate
if a similar halo component is indeed present in NGC 3109.
Hereafter we define Pop II as the stellar component that is
very old ($log~t >10$ yr), and metal-poor ($Z<0.002$, or $Z/Z_{\odot}<0.1$)
(see the reviews of Hodge 1989 and Mateo 1998 for a discussion of star
formation histories and Pop II stars in the Local Group galaxies).

In this paper we present deep optical photometry covering a large
field centered on NGC~3109, which allows
a detailed photometric study of individual stars and stellar populations
in this galaxy.
The observations of NGC~3109, data reductions, and photometry are described in
Section 2 and the resulting color-magnitude diagrams are given in
Section 3, along with a discussion about reddening. 
The spatial distribution of the different stellar populations is discussed
in Section 4.  Section 5 presents the luminosity function of NGC~3109.
Fundamental parameters of NGC~3109, such as metallicity, age and distance,
are determined in Section 6. Section 7 discusses other population
tracers, namely star clusters, carbon stars, HII regions, planetary nebulae, 
Cepheid variables, CO and HI observations.
The formation and evolution of NGC~3109 are discussed in Section
8 and the conclusions of this work are summarized in Section 9.

\section {The Data}

\subsection {Observations and Reductions}

The observations of NGC~3109 were obtained during the night of January 7,
1995, as part of a long term monitoring program of the variable stars
in several galaxies (Zijlstra et al. 1996a). 
We used the red arm (RILD mode)
of the ESO Multi-Mode Instrument (EMMI) at the New Technology
telescope (NTT) in La Silla, operated remotely from the
ESO headquarters in Garching, Germany (see Zijlstra et al. 1997).  
The weather was photometric, with seeing varying from $0.8\arcsec$ to
$1.1\arcsec$ during the night.
We used the 2048$\times$2048 Tek
CCD, with a scale of $0.268\arcsec$ pix$^{-1}$.  

The observations of the NGC~3109 galaxy were secured with the 
Johnson $V$ and Cousins $I$
filters. They consist of pairs of frames taken at mean airmass 1.1,
with exposure times 900 sec
in $I$, and 1800 sec in $V$. These frames, 
shown in Figure 1, are
centered on the galaxy coordinates listed in Table 1.
A total of 5 standards of Landolt (1992)
were observed during the night in order to calibrate our
photometry.

The reductions of the CCD frames were carried out following standard
procedures, using the package CCDRED within the IRAF environment. 
The pairs of images in each filter were combined, the final FWHM
of stellar images being 0.9 and 0.8 arcsec in V and I, respectively.

An additional
pair of 900-sec $I$-band frames (airmass 1.25) was obtained $8\arcmin$ South
of the center of the galaxy during the night of April 16, 1996, 
in order to account for field contamination and study 
the possible presence of an extended halo around this galaxy.  
These $I$-band images were processed in similar way as the previous data.  
The location of the background
frame in the sky is directly to the South of the main NGC~3109 field shown
in Figure 1.
The field to the S of NGC~3109 is rich in background galaxies.  

The fields cover most of the light of the galaxy included within the outer
contours $\mu_B= 25.0$ mag arcsec$^{-2}$
of the surface photometry of Jobin \& Carignan (1990), and 
our deep photometry  extends beyond the
Holmberg radius ($R_H = 2.8$ arcmin).
The total area covered is $204 ~arcmin^2$.

The photometric transformations to the standard system
were done following the procedure described
by Minniti \& Zijlstra (1997).
These transformations reproduce the 
magnitudes of the standard stars with an $rms$ of 0.016 in V and 0.031 in I.

The photometric measurements were performed within the IRAF
environment, using DAOPHOT II which is an improved version of the
original DAOPHOT package developed by Stetson (1987).  In particular,
DAOPHOT II can accommodate a spatially varying point-spread function
across the field. This is needed for our frames, which cover a large
field.  All stars in the $V$ and $I$ frames of NGC~3109 with more 
than 5$\sigma$ above the background were located, and their magnitudes
measured by fitting a Moffat point-spread function. The resulting
limiting magnitudes ($5\sigma$) are V$= 25$, and I$= 24$.

As expected, the completeness is worse in the more crowded disk region of
NGC~3109 ($C \sim 75\%$) than in the outer regions ($C \sim 90\%$).
However, the present
photometry is sufficiently deep that none of our results will depend
on the completeness of the sample at the faintest magnitudes.

NGC 3109 is located at a relatively low Galactic latitude
($l=262.1^{\circ}, ~b=-23.1^{\circ}$), and contamination
by foreground stars becomes an issue.
The foreground stars from the Galactic halo and disk should appear as a plume
of stars in the color-magnitude diagram with mean color $V-I = 0.7$, and
a total color range covering about $0.4 \leq
V-I \leq 1.5$.  The contamination from these stars can be estimated
from star counts in a strip along declination of $0.5 \times 8.9$
arcmin, offset by 10 arcmin from the NGC~3109 major axis. The  observed
density of red stars
in this region is $<$1 stars/arcmin$^2$ with $I\leq 20$, and 4
stars/arcmin$^2$ with $20.5\leq I \leq 22.5$. 
{}From the Galactic model of
Ratnatunga \& Bahcall (1985), we expect 0.7 foreground halo
stars/arcmin$^2$ with $V \leq 21$, and $0 \leq V-I \leq 2$ in this
low-latitude field.  

Since the galaxy/star ratio increases rapidly for faint magnitudes,
the background galaxy contamination also has to be  taken into account.
Most of
the brighter galaxies are resolved and are discarded by our sharpness
criterion; only a few fainter and very compact ones may cause
confusion. {}From Tyson (1988) we expect a galaxy density of 
$10^4 gal/degree^2~mag$ with
$I<21.5$. This would give $\sim 227$ galaxies in our CCD field.
Most of these galaxies are
resolved, and were eliminated on that basis.

\section{The Color-Magnitude Diagrams}

Optical color-magnitude diagrams have been the primary tools to study the
past history of star-formation in dwarf irregular galaxies of the
Local Group. (e.g. Lee 1993, Lee et al. 1993, Lee 1995,
Tosi 1994, Tolstoy 1995, 1996, Marconi et al. 1995, 
Gallart et al. 1996, Aparicio et al. 1997a, 1997b, Dohm-Palmer et al. 1997).

Comparable CCD photometry of smaller areas around NGC~3109 has been published:
BVRI photometry of Bresolin et al. (1993), BVI photometry of 
Davidge (1993), VR photometry of Greggio et al. (1993), and
VI photometry of Lee (1993).
The present report gives photometry for a larger
number of stars than these previous studies, having a larger
spatial coverage, which allows us to study in detail the stellar
populations and their spatial distribution across the face of NGC~3109.
Star-by-star comparisons with these papers are not possible, but
we note that in the overlap regions, these published
color-magnitude diagrams are very similar to ours.
Infrared photometry of a small field in the central region of this
galaxy are presented by Alonso et al. (1998).
In addition, photographic photometry also exists in the literature:
BV photometry of Sandage \& Carlson (1988), and of
Demers et al. (1985).

Figure 2 shows the $I ~vs ~V-I$, and $V ~vs ~V-I$
color-magnitude diagrams for a total of 15000 stars in NGC~3109
with centroids matched in $V$ and $I$ frames to better than 2.0 pixels
(0.6"), and satisfying stringent criteria of photometric quality
($\sigma \leq 0.5$, $\chi \leq 2$, and sharpness $\geq -1$). The
sharpness criterion eliminates galaxies as well as some possible star
clusters.  The main features of these color-magnitude diagrams are
a red tail of asymptotic giant branch (AGB) stars,
a blue main sequence with mean color $V-I = -0.2$, and a 
red supergiant sequence with $I < 20.5$, reaching $I \approx 17.5$. Signs
of a population of blue loop stars are also seen, running  in a sequence
about 0.3 mag redder, parallel to
the main sequence. The AGB is very extended, and can be traced to
$V \approx 25$, and $V-I \approx 3.6$. Because of the low metallicity,
this red color AGB may be dominated by carbon stars. 
Note that Galactic foreground stars have not been removed from this diagram.

The complexity of the stellar content of NGC~3109 is revealed when considering 
the color-magnitude diagrams of different regions.  Greggio et al. (1993) 
point out that the star formation history differs in
different NGC~3109 disk fields. They find a metallicity $0.001<Z<0.01$ based on
VR photometry, and conclude that the star formation was more or less continuous 
during the past Gyr.
Figure 3 shows the $I$ $vs$ $V-I$ color-magnitude diagram 
of NGC~3109 compared with the 
theoretical isochrones of Bertelli et al. (1994)
for $Z=0.001$ ($Z/Z_{\odot}=0.05$) and $Z=0.004$ ($Z/Z_{\odot}=0.2$).
The isochrones with ages $log~t= 6.6, 7.0, 7.8, 8.0, 8.7, 9.0, 9.7$ 
and $10.0$ years
have been shifted according to  the distance and reddening listed
in Table 1. Throughout this work, the isochrones are plotted as
discrete points in order to illustrate the varying timescales of evolution.
It is clear that stars of different ages are present in Figure 3,
as well as different abundances. By comparison with the color-magnitude
diagrams of similar galaxies (the LMC Shapley Constellation
III from Reid et al. (1987), the SMC from Reid \& Mould (1990), and WLM
from Minniti \& Zijlstra (1997)), 
and with the theoretical isochrones, the stellar disk
of NGC~3109 seems to have a mean metallicity that is intermediate
between $Z=0.001$ and $0.004$.

\subsection{Reddening}

The reddening towards NGC~3109 is thought to be small.  Based on the HI
column density, Burstein \& Heiles (1984) estimate the foreground
reddening as $E(B-V)=0.04$. Schlegel et al. (1998) obtain a similar
value based on the COBE and IRAS  maps at 100$\mu m$.
This is  equivalent to $E(V-I) = 0.05$
and $A_I = 0.08$ using the reddening ratios $E(V-I) = 1.60 ~E(B-V)$, 
$A_I = 1.49~ E(B-V)$ of Rieke \& Lebofski (1985). 
Based on optical
photometry, Lee (1993) adopts $E(B-V) = 0.04$, and Sandage \& Carlson (1988) adopt
$E(B-V) = 0.00$. Davidge estimates a total reddening (internal $+$ foreground)
of $E(B-V) = 0.14$ for a field centered on the disk.  

We can estimate the foreground reddening by using the sharp blue cutoff in
the stellar distribution of the outer NGC~3109 fields (the outer regions are
used to avoid internal disk reddening). 
This cutoff is interpreted as the
locus of the main sequence, which is nearly vertical in the
theoretical color-magnitude diagrams, having an intrinsic color of
$(V-I)_0 = -0.20$ (Bertelli et al. 1994).  {}From the observed color
we deduce $E(V-I) = 0.03$, consistent with the $op.~cit.$ value.
Given these considerations, we adopt  
$E(V-I) = 0.05$, and $A_I = 0.08$ for NGC~3109.  

We also investigate the presence of differential absorption across
the face of NGC~3109. 
The $V-I$ color of
the RGB and the blue main sequence (MS) on both sides (E and W) of NGC~3109 differ at the
$0.10$ mag level in the mean. 
The obscuration is less
severe in the W arm. However, localized regions of higher obscuration are present
inside the disk.

\section{Spatial Distribution of Stars: the Halo of NGC~3109}

Figure 4 shows the
color-magnitude diagrams of an NGC~3109 halo field 
compared with an inner disk field.
The differences are striking, although some effects must be taken
into consideration. First, the disk regions are more
crowded, with the completeness and photometric errors worse than
in the outer regions. Second, we should also expect more differential reddening
within the disk than in the halo fields. Third, the proportion of blue to
red stars is much larger in the disk field than in the halo fields.
However, there is no way in which crowding or reddening
can transform the halo color-magnitude diagram (left panel) 
to make it look like that of the NGC~3109 disk (right panel). 
The population differences are significant.
Figure 4 is similar to the color-magnitude diagrams of the central
$3.8\times 3.9$ arcmin$^2$ of Lee (1993), confirming that
the population differences between disk and halo extend to a larger 
area and to both sides of the plane of this galaxy.

Figure 5 shows the same color-magnitude diagrams as Figure 4, with the
isochrones of Bertelli et al. (1994) corresponding to a 
metallicity  $Z=0.001$, and ages 
$log~t= 6.6, 7.0, 7.8, 8.0, 8.7, 9.0, 9.7$
and $10.0$ years. It is clear that a wide range of ages is present in the
disk field, while the halo field contains only an old population.
In order to emphasize this,
Figure 6 shows the color-magnitude diagram of the NGC~3109 halo field ,
zooming in the RGB region,
compared with a similar diagram of the halo field in the dwarf irregular
galaxy WLM, studied by Minniti \& Zijlstra (1996, 1997). These diagrams are
similar, with the NGC~3109 giant branch being slightly bluer, indicating a 
slightly more metal-poor population. Also, the FWHM of the NGC~3109 giant
branch is similar to that of WLM, in spite of the fact that the giant stars
in NGC~3109 are nearly a magnitude fainter than the giant stars in WLM
of identical intrinsic luminosity.

The spatial distribution of the blue and red stars are quite
different: Figure 7 shows the dependence of $V$ and
$V-I$ as function of projected height above the NGC~3109 plane $z$ in pixels.
The radial extent of the present observations reaches 
three times as far ($11.5\arcmin$) as the outer contours ($3.5\arcmin$) of the
surface photometry of Jobin \& Carignan (1990) that correspond to a surface
brightness $\mu_B=25.0$ mag arcsec$^{-2}$. 

As shown in Figure 7, most of the stars detected in the V frames down
to $V = 24.5$ are concentrated in the disk of the
galaxy, within $\sim$2.5 arcmin of the major axis (equivalent to
$z=1$ kpc projected distance from the plane of this galaxy). However, there is a
low density stellar component that extends as far as the edge of our field,
out to $\sim$4.5 arcmin (or $z=1.8$ kpc from the NGC~3109 plane).
Hereafter, $z$ refers to the position measured in pixels along the
direction of declination with respect to the plane of NGC~3109.
With the scale of $0.268$ arcsec pix$^{-1}$, 
$1$ arcmin is equivalent to $0.4$ kpc at the distance of $D=1.33$ Mpc. 
This galaxy is reasonably edge on, with an inclination 
$i=80^{\circ}\pm 2^{\circ}$ (Jobin \& Carignan 1990).
Seeing the galaxy edge-on makes the study of  the disk
population more difficult, because there is no depth information:
different regions (with various metallicities, 
star formation histories and reddenings) along the line of sight overlap.
At the same time, however, two advantages are clear
because of NGC~3109 being nearly edge-on:
1. the separation between a halo and a disk population is simplified, and
2. disk reddening will not affect the halo population.

The radial dependence of $V-I$ is also shown in Figure 7. 
The red stars with $1.0 < V-I < 1.7$
are less concentrated than the blue stars with $V-I \leq 0.5$, 
extending as far out as $z=4.5 \arcmin$.  The appearance of this
diagram can be understood in terms of a superposition of an extended old and
metal-poor population, plus a younger, more concentrated component,
where scatter may be introduced by possible differential reddening and
by star crowding in the inner regions.

The HI extends almost
as far out as the NGC~3109 halo, to $z=3.5$ arcmin, as shown by the HI maps
(Jobin \& Carignan 1990).
There is evidence of a warp in the disk of this galaxy from these HI maps.
We will assume that the extended component belongs
to the halo and not to a highly warped disk, because the populations are
too different. If the old metal poor component observed away from the
projected plane of NGC~3109 were in the warp, the disk itself must have
a strong population gradient.

We cannot measure the flattening of the NGC~3109
halo, because this halo clearly extends
outside of our fields in the E-W direction. 
Only an upper limit to the flattening can be derived, $b/a < 0.6$, given
by the fact that the total disk extent is at least 11 arcmin 
(Demers et al. 1985). If the halo flattening were the same as the disk 
seen in projection ($b/a = 0.2$), this halo must have an extension of
about $40$ arcmin in the E-W direction. For a rounder halo, with
$b/a = 0.5$, this extension would be about $16$ arcmin. 
Note that the halos of the large spirals of the Local Group, MW and
M31, are also flattened. Larsen \& Humphreys (1994) measured an axial
ratio of $0.60\pm 0.05$ for the MW halo, and Pritchet \& van den Bergh
(1994) measured an axial ratio of $0.55\pm 0.05$ for the M31 halo.
We also estimated that the halo of the dIr galaxy WLM is significantly
flattened, with an axial ratio of $0.6\pm 0.1$ (Minniti \& Zijlstra 1997).

\section{The Luminosity Functions}

The luminosity function was constructed by counting all the stars
present in the I-frames ($N=17500$), regardless of them matching the V-images.
The normalized counts of stars are
plotted in Figure 8. This luminosity function is dominated by the disk stars,
and is similar to that of the WLM disk (see Minniti \& Zijlstra 1997).

The luminosity function for the NGC~3109 halo field is shown in Figure 9.  
Two distinct features can be seen in the luminosity function:
a sharp break at $I _{RGBT}= 21.7\pm 0.05$ due to the
termination of the halo-like RGB, and another break at $I_{AGBT} = 20.5$ due
to the termination of the AGB. Lee (1993) found
the RGB break  at $I_{RGBT}= 21.55\pm 0.1$, in reasonable agreement with our
measurement.  The signature for the RGB termination at $I_{RGBT} = 21.7\pm 0.05$
is clearer in the halo, and it is seen in the halo luminosity functions 
as far away as $z=4$ arcmin from the NGC~3109 plane.  This RGB tip 
allows us to measure an accurate distance to NGC 3109, as discussed 
in the next section.

We also construct a bolometric luminosity function for the red giant and
supergiant stars with $V-I>0.7$ using
$$M_{bol}=I_0+0.30+0.38(V-I)_0-0.14(V-I)_0^2-(m-M)_0$$
(Brewer et al. 1995).
This bolometric luminosity function is shown in Figure 10.
The brightest red supergiants that are likely to be members of NGC~3109 reach
$M_{bol}=-9.5$. 
Note that the brightest LMC supergiant is WOH G064 (IRAS04553-6825),
with $M_{bol}=-9.3$ (Chiosi \& Maeder 1986, Zijlstra et al. 1996b).
There is also a break (increase in the number of stars)
at about $M_{bol}=-6$, corresponding to a numerous $\sim$1 Gyr old population
like in the LMC. Brighter stars, with $M_{bol}=-7.5$, can either be AGB
stars which temporarily experience hot bottom burning, or supergiants.

\section{Fundamental parameters of NGC~3109}

\subsection{Metal Abundance for the Halo Population}

The metallicity of old and metal-poor populations, such as we find in
the halo around NGC~3109, can be derived from the V$-$I color of their
RGB. Da Costa \& Armandroff (1990) have shown that for halo globular
clusters, this parameter is very sensitive to metallicity.
We use the recent calibration of the V$-$I color versus abundances
of Lee (1993). This calibration is based on the metallicity
dependence of the $V-I$ color of the RGB  0.5 mag below the tip,
at $M_I = -3.5$.  Figures 3 and 4
show that $V-I = 1.35$ at $M_I=-3.5$ ($I=22.2$) for the halo of NGC~3109, 
from which we derive $[Fe/H] = -1.8\pm 0.2$ using the calibration of Lee (1993).
Using a similar approach, Lee (1993) measured $[Fe/H] = -1.6\pm 0.2$,
and Davidge (1993) estimated $[Fe/H] < -1.6$, in excellent agreement with our
result.

We use the color spread of the RGB at a fixed magnitude  in order 
to study the range of metallicities in the stellar population of NGC~3109.
The mean color of the
RGB becomes redder as the metallicity increases.
For example, the difference between the isochrones with $[Fe/H] = -2.0
$ and $-1.0$ is $\Delta (V-I) = 0.30$ mag at $M_I=-3.5$.
We use the colors of the RGB at 0.5 mag below the RGB tip ($M_I=-3.5$)
because the photometric errors increase going to fainter
magnitudes.  The spread in the V--I color at
$I=22.2$ is $\Delta V-I = 0.30$, implying a metallicity range
of $\Delta [Fe/H] \approx 0.3$ dex. The photometric uncertainty at this
magnitude level is $\sigma_{V-I} = 0.05-0.07$, not affecting this estimate
of the metallicity spread.

In summary, the halo of NGC~3109 has $[Fe/H] = -1.8 \pm 0.2$ and the
metallicity range is $\Delta [Fe/H] \approx 0.3$ dex.  
The abundances are comparable to those of the dwarf irregular
galaxy WLM (Minniti \& Zijlstra 1997).

\subsection{The Distance}

Lee et al. (1993) improved the method for obtaining accurate
distances for distant galaxies from VI photometry of their brightest
stars.  They demonstrated that this method is
perhaps as accurate as other primary distance indicators such as RR
Lyrae and Cepheid variables. This method is  based on the fact that
$M_I = -4.0$ for the red giant branch tip of
metal-poor stellar populations with $[Fe/H] < -1.0$.

The halo of NGC~3109 is very metal-poor, as discussed above,
where we have determined $[Fe/H] = -1.8\pm 0.2$.
Thus, we can apply the RGB tip to measure the distance of this galaxy.
The RGB tip of the NGC~3109 halo is located at $I=21.7\pm 0.05$ (Figure 7).
For $A_I = 0.08$, the
distance modulus to NGC~3109 is $(m-M)_0 = 25.62 \pm 0.1$, which translates to  
$D = 1.33$ Mpc.
Because we are using the halo stars on NGC~3109, this distance is
not affected by internal extinction.
The distance measured here
is in excellent agreement with the recent determinations of
Lee (1993) and  of Musella et al. (1997).

\subsection{The Ages}

We can obtain a rough age estimate by measuring
the magnitude difference between the tip of the RGB and the tip of the AGB, $\Delta TT$
(Minniti \& Zijlstra 1997). From
the isochrones of Bertelli et al. (1994), there is a well-defined
relationship between $\Delta TT$ and age at any given metal
abundance. 
The pure halo fields of NGC~3109 are characterized
by a single mean abundance, allowing us to apply
this method.  For a metallicity of $Z=0.0004$ to $Z=0.001$ (and $Y=0.23$), the
relationship between $\Delta TT$ and age 
computed using the isochrones of Bertelli et al. (1994)
gives an absolute age of $t = 10^{10}$ yr for
$\Delta TT = 0.8-0.9$ mag.

This estimate is independent of
reddening and of the distance to NGC~3109, and agrees with the isochrone fits.
However,
this method is based on isochrones which may be affected by
systematics (e.g. uncertainties in mass loss, convection, opacities,etc).  
An alternative way to reliably detect the presence of an
old population, independent of isochrones, is to measure the jump of
the luminosity at the tip of the RGB. As shown by Renzini (1992), an increase
of more than a factor of 4 in this break indicates
the existence of old stars. This is indeed the case of NGC~3109, as seen after
we substract the background contamination from the luminosity function
plotted in Figure 6.

These approximate age determinations, added to the presence of  globular 
clusters in the NGC~3109 halo, argue in favor of an
old age ($10^{10}$ yr) for the NGC~3109 halo. 
However, confirmation of the old age is needed, such as possible with 
deep HST photometry of the horizontal branch.

\section{Other Population Tracers}

\subsection{Star Clusters in NGC~3109}

The status of globular cluster research in galaxies of the Local Group
is summarized by Olszewski (1994).  NGC~3109 has 10 globular
cluster candidates located outside the disk (Demers et al. 1985),
none of which are confirmed spectroscopically.
These clusters are fainter than typical Milky Way globular clusters.
Unfortunately, none of these clusters are located in the fields studied 
here. In the inner crowded regions of this galaxy, however,
some other clusters may be hiding, as NGC~3109 is similar to
the Magellanic Clouds in many respects. 

\subsection{Carbon Stars}

The C star luminosity functions given by Brewer et al. (1995) for M31,
and by Richer (1981) for the LMC bar West field peak at $I_0 =
19.6$, with a total range of $18.8 \leq I_0 \leq 20.5$.
This is coincident with the locus of the red tail seen in
Figure 2. These stars are located within the main body of the galaxy.

\subsection{HII Regions}

Hodge (1969) found that NGC~3109 contains several HII regions.
These HII regions are mostly minor, as NGC~3109 is not very 
active in forming stars. 
Ten HII regions identified from [OIII]$\lambda 5007$ images
have been studied  recently by Richer \& McCall (1992).
Figure 11 shows the color magnitude diagrams of the regions surrounding
these HII regions. They were made using fields with $100$ pix on a side,
or $0.2$ arcmin$^2$. The panels of Figure 11 give a panoramic view on
the ages of the stars in the HII region fields. 
Using the brightest main sequence stars we can set limits on
the ages of the youngest stars in these regions. Table 2 lists 
the positions of the shell centers in our images, the fields covered, 
the age limits derived from the blue main sequence stars and red blue-loop
stars, and the total numbers of stars plotted in Figure 11.
In all cases, there clearly are very young stars present in
the fields of the HII regions.  We note that in these regions of the
galaxy, there is unresolved background. We also see different blue
edge colors for the main sequence stars, evidence of different amounts of
obscuration.  The ages measured for these regions are consistent with 
some star formation in the recent past, that has been more or less continuous,
as discussed by Greggio et al. (1993).

Richer \& McCall (1995) measure a low metallicity
$12+log(O/H)=8.06$, equivalent to
about $Z=0.001$, similar to the SMC and WLM. Within the errors,
this abundance is consistent
with the stellar abundance derived here, suggesting
moderately low star-formation rate in the past.
Since the gas reflects the
cumulative process of enrichment along the life of a galaxy, in the
absence of infall one is led to conclude that the past star-formation
activity was not substantial enough to raise this composition significantly. 
Thus, the metallicities of the disk stars inferred from the color-magnitude 
diagrams are in agreement with the gas compositions.

Shells and supershells have been found in NGC~3109 from H$\alpha$ surveys
(e.g. Hunter et al. 1993), revealing a complex interstellar medium and
active star formation. The panels of Figure 12 show the color-magnitude
diagrams of the regions of supershells 3 and 4, and shells 2, 5, 6, 7, 8, 8, 11,
and 12. Hunter et al. (1993) pointed out that not all of these shells
are surrounding bright blue stars.  We can quantify this statement using
the deep VI photometry. The color-magnitude diagrams shown in Figure 12
cover fields with 100 pixels on a side ($0.2$ arcmin$^2$) centered
in these shells.
As with the HII regions, using the brightest main sequence stars 
we can set limits on the ages of the youngest stars in these regions. 
These age limits are also listed in Table 2, along with the central
positions of the shells, the field covered, and the total number of stars
plotted.  We caution that these
panels show all the stars projected along the line of sight, some
of which may be foreground or background to the shells, but still within the
NGC~3109 disk. Most of these shells surround groups of very young stars.
However, it is clear that not all these regions contain very young stars
that would drive the gaseous material away. In particular, the color-magnitude
diagrams of shells 6, 11 and 12 are dominated by old stars. Shells 11 and
12 are located above the plane of NGC~3109, explaining the smaller 
number of stars in their color-magnitude diagrams.

\subsection{Planetary Nebulae}

Richer \& McCall (1992) found 7 planetary nebulae in NGC~3109 
based on $[OIII] \lambda 5007$ \AA\ images. They
used their luminosity distribution 
to estimate a distance to this galaxy, concluding that
$(m-M)_0=25.96$. Even though this value is larger 
than the present estimate, their quoted error bars of about 0.5 mag makes
this distance also consistent with our results.

During an NTT observing run in December 1996 we obtained a 900 second 
exposure centered on NGC~3109 with the SII narrow-band filter ($\lambda 9500\AA$).
Because it is redder than $[OIII] \lambda 5007\AA$,
this filter is well suited to find obscured planetary nebulae. 
Unfortunately, there is a shift of wavelength sensitivity with position
in the detector that is not well mapped, and our search area may be
smaller than the actual $9\times 9$ arcmin observed. In any case, 
we have not detected any other bright PN in the central regions
of NGC~3109.

\subsection{Cepheid Variables}

Sandage \& Carlson (1988) discovered 29 Cepheids in NGC~3109, and 
used them to measure the distance of
this galaxy.  The most recent study by Musella et al. (1997)
reports BVRI photometry for 36 candidate Cepheids.
They use the LMC Cepheid period-luminosity relation to obtain a relative
distance modulus of $\Delta \mu = -7.10$ mag between the LMC and NGC~3109.
This corresponds to $m-M_{N3109}=25.67$, for $m-M_{LMC}=18.50$.
This distance is in excellent agreement with our result.
Our ongoing monitoring project (Zijlstra et al. 1996a)
will detect all the Cepheids down to $I = 23$ in this galaxy.

\subsection{The Gaseous Component: HI Observations}

Dwarf irregulars are dominated by dark matter halos, their total M/L being larger
than for normal spiral galaxies (Jobin \& Carignan 1990, C\^ot\'e 1995).
The most recent HI maps of NGC~3109 are given by Jobin \& Carignan (1990),
who mapped the whole galaxy at high resolution and sensitivity.
The HI emission is detected out to 3.5-4.0 arcmin from the NGC 3109 disk,
almost as far as the stars in the halo component found here.
Jobin \& Carignan (1990) also give a 
detailed HI rotation curve for NGC~3109 based on these observations.
The amplitude of this curve is large for such
a small galaxy ($\sim 50$ km/s), implying a high M/L.
Indeed, NGC~3109 is taken as a typical dwarf irregular galaxy with a
sizeable dark matter halo (Navarro et al. 1997).

\section{Discussion: Formation and Evolution of NGC~3109}

By analogy with other spirals of the Local Group,
we define halo as an old and metal-poor population of stars that are distributed
in a spheroid. Such a population would be kinematically hot, with small rotation
and large velocity dispersion. More quantitatively, a halo
of Pop II stars is defined here as $\geq 10^{10}$ yr old and metal-poor
$[Fe/H] \leq -1$ stars (see the reviews by Hodge 1989 and Mateo 1998). 

Mould \& Kristian (1986) established the presence of Pop II halos in M33 and
M31 based on the RGB morphology, before there was kinematic
information about these halos. Later confirmation that these
halos were kinematically hot came for example from radial velocities of 
globular clusters (Schommer et al. 1992, Brodie \& Huchra 1991).   

The present observations show that NGC~3109 has an old and metal-poor
population of stars that is more extended
than the younger, bluer stars that are located within the disk of the galaxy.
While  the evidence suggests that this population forms a halo such as
seen in Local Group spirals, this requires confirmation from
kinematics, which can show if this population does not participate in
the rotation of the disk. Useful targets for radial velocity measurements
would be the candidate clusters of Demers et al. (1985), and individual
giants in this galaxy.

All well studied galaxies of the Local Group,
regardless of type and luminosity, seem to have population II stars. This
conclusion was based on the presence of globular clusters, RR Lyrae
variables, or blue horizontal branches and metal-poor giants
(Hodge 1989, Da Costa 1994, Grebel 1998, Mateo 1998). 
In the case of NGC~3109, the only evidence was given by the presence of
several candidate globular clusters.
The present work strengthens the evidence of an old and metal-poor halo in
this galaxy.
Note that the presence of a halo population around dIr galaxies may be common:
Minniti \& Zijlstra (1996, 1997), and Aparicio et al.  (1997b)
have found such a population in the dIr galaxies WLM and Antlia, respectively.

As discussed in the Introduction,
the existence of such halos would be very significant in the 
cosmological context, since it may be related to 
dark matter halos, if the dark matter is baryonic (Alcock et al. 1997).
A population of low mass pre-galactic stars would be essentially
collisionless (Miralda-Escud\'e \& Rees 1997, Loeb 1997), 
populating the diffuse halos of galaxies such as NGC~3109. 
The degree to which the massive compact objects detected through microlensing
are related to the old halos observed in dwarf irregulars like WLM and NGC~3109
is heavily dependent on the initial mass functions (e.g. Chabrier et al. 1996). 
Unfortunately, these
galaxies are distant enough to prevent observations of their old main sequence.
The only information available so far comes from evolved stars, and their 
initial mass functions remain unconstrained. The old main sequence is
within reach for galaxies out to about 1 Mpc, using STIS at
the Hubble Space Telescope (Gregg \& Minniti 1997).

The existence of a Pop II halo in dwarf irregular galaxies such as NGC~3109
and WLM is also important in the context of galaxy formation.
NGC~3109 and WLM  give examples of very small mass  galaxies that have formed
halos of their own.  Because they have a small mass, merging of smaller subunits
or strong interactions seem less likely than in the Milky Way, for example
(particularly in the case of WLM).
In our galaxy, the formation of the halo may have occured in part 
by accretion of 
smaller satellites, as predicted by the CDM clustering model (e.g. White 1996).
These two galaxies formed at the edge of the Local Group but in opposite sides. 
Indeed, Peebles (1995) finds that
NGC~3109, like WLM, has turned around from the Hubble flow and is free 
falling towards the Local Group center of mass.
The recently discovered Antlia dwarf is the nearest known companion to
NGC~3109, at a projected separation $\Delta r=30$ kpc (Aparicio et al. 1997b).
Their relative velocity is $\Delta V=45$ km s$^{-1}$ (Aparicio et al. 1997b),
much smaller than the typical velocity differences of Local
Group galaxies ($\Delta V = 400$ km s$^{-1}$), 
suggesting that these galaxies may be associated. We speculate that
at this relative velocity they may have
suffered a close encounter within the last Gyr, triggering or sustaining
the star formation.

\section{Conclusions}

We have obtained deep VI photometry of about 17500
stars in the field of the dwarf
irregular galaxy NGC~3109, located at the edge of the Local Group. 

We study the structure of NGC~3109, in particular examining
the spatial distribution of blue (young) and red (older) stars.
The main result of this work is to establish 
the presence of an extended stellar halo, down to very faint
levels. The stars populating this halo are clearly different
from the stars that make the NGC~3109 disk: they are old and metal-poor 
(Figure 3).

{}From the VI color-magnitude diagram we infer metallicities 
and ages 
for the stellar populations in the main body and in the halo of NGC~3109.
The color-magnitude diagrams also allow us to confirm a low reddening
for this galaxy, $E(B-V) = 0.04$.

We give an approximate determination of the age of the NGC~3109 halo, independent
of the reddening and distance to this galaxy, concluding that it is very
old ($log~t \sim 10$ yr). 
The age measurement depends on
the metallicity, which is accurately determined from the color of the
RGB, and which agrees with previous determinations in the main body of
the galaxy (Richer  \& McCall 1995).
This halo is also metal-poor, with $[Fe/H]=-1.8\pm 0.2$, characteristic 
of Pop II stars.

We construct deep optical and bolometric
luminosity functions, obtaining an accurate distance for
this galaxy $(m-M)_0=25.62\pm 0.10$, based on the RGB tip of the metal poor 
stellar component.  This value is in excellent agreement with recent
distance determinations using Cepheids (Musella et al. 1997).

The real significance of dIr galaxies as possible elements contributing
to build larger galaxies like our own has been overlooked, perhaps under
the assumption that, like the LMC, they do not have Pop II halos. 
However, we find that 
dIr galaxies can have halos. NGC~3109 represents another such case
after WLM (Minniti \& Zijlstra 1996), where no spiral
arms, nucleus or bulge are present, but where a disk formed dissipatively 
within an old and metal-poor halo. 

Measuring the kinematics of the NGC~3109 disk and halo components
using radial velocities of individual giants is the necessary next step.
This would be within the reach of current instrumentation:
the ESO Very Large Telescope equipped with FORS
would be ideally suited for getting accurate enough
radial velocities of large numbers of individual NGC~3109 giants.
These observations would unambiguosly determine if the NGC~3109 halo is
kinematically hotter (with smaller rotation and larger velocity
dispersion) than the disk, as we would predict based on the present work.

\acknowledgements{
We are grateful with J. Rodriguez, V.
Reyes and M. Pizarro for their help at the NTT.
This work was performed in part under the auspices of the U. S. Department of 
Energy by Lawrence Livermore National Laboratory under Contract W-7405-Eng-48.
MVA thanks CONICOR, CONICET and Fundacion Antorchas of Argentina for finantial
support.}

\clearpage
\begin{figure}
\caption{
CCD $9\arcmin \times 9\arcmin$ field of NGC~3109 observed with the NTT.
North is up, and East to the left. At the distance of NGC~3109,
$1\arcmin = 0.39 ~kpc$. }
\end{figure}

\begin{figure}
\caption{
Color-magnitude diagrams for all the stars with $V$ and $I$ measured in NGC~3109.
}
\end{figure}

\begin{figure}
\caption{
VI color-magnitude diagrams of NGC~3109 compared with theoretical
isochrones of Bertelli et al. (1994)
for $Z=0.001$ (left panel) and $Z=0.004$ (right panel).
The isochrones have been shifted according to  $(m-M)_0=25.62$ and
$E_{V-I}=0.05$ measured in this work.
>From top to bottom, the isochrones shown have ages $log~t= 6.6, 7.0, 7.8, 8.0, 8.7, 9.0, 9.7$
and $10.0$ years.
}
\end{figure}

\begin{figure}
\caption{
Color-magnitude diagrams of an NGC~3109 halo field -- left panel --  
compared with an inner disk field -- right panel -- (see text).
}
\end{figure}

\begin{figure}
\caption{
Color-magnitude diagrams of the NGC~3109 halo  -- left panel -- 
and disk fields  -- right panel -- compared 
with the Bertelli et al. (1994) isochrones for $Z=0.001$.
The isochrones have been shifted according to  $(m-M)_0=25.62$ and
$E_{V-I}=0.05$ measured in this work.
>From top to bottom, the isochrones shown have ages $log~t= 6.6, 7.0, 7.8, 8.0, 8.7, 9.0, 9.7$
and $10.0$ years.
This figure shows that there are stars of all ages in the disk, while the
halo is made only of old stars.
}
\end{figure}

\begin{figure}
\caption{
Color-magnitude diagram of the NGC~3109 halo field compared with the WLM halo field.
Note the tightness of the red giant branch, with the tip  located
at $I=21.7$. The NGC~3109 red giant branch is slightly bluer than that
of WLM, indicating a lower metallicity
(see text).
}
\end{figure}

\begin{figure}
\caption{
Magnitudes and colors of individual stars as function of projected
height $z$ above the NGC~3109 disk. This figure shows that the
young stars (i.e. the brightest stars in the upper panel, and the
bluest stars in the lower panel) are concentrated in the disk.
A population of old stars (faint and red) is also clearly seen,
extending at least out to the edges of the field.
}
\end{figure}

\begin{figure}
\caption{
Luminosity function of NGC~3109, including all the stars 
detected in the disk and halo fields.
Beyond $I=22$ the counts in the inner regions are incomplete.
}
\end{figure}

\begin{figure}
\caption{
Luminosity function of the NGC~3109 halo field.
Note the rapid increase in number counts due to the RGB termination 
at $I = 21.7$. The AGB termination is also seen at $I = 20.5$.
Incompleteness in these regions causes the luminosity function
to change slope beyond $I=23$.
}
\end{figure}

\begin{figure}
\caption{
Bolometric Luminosity function of NGC~3109, including all the red
stars with $V-I>0.7$.
Note the rapid increase in number counts due to the AGB termination 
at $M_{bol} = -5.8$.
The counts are incomplete for $M_{bol}>-3.25$.
}
\end{figure}

\begin{figure}
\caption{
$V$ vs $V-I$ color-magnitude diagrams centered in the HII regions studied by
Richer \& McCall (1992),
compared with theoretical isochrones of Bertelli et al. (1994)
for $Z=0.001$.  The isochrones with ages $log~t= 6.6, 7.0, 7.8, 8.0, 8.7, 9.0, 9.7$
and $10.0$ years have been shifted according to  $(m-M)_0=25.62$ and
$E_{V-I}=0.05$ measured in this work.
}
\end{figure}

\begin{figure}
\caption{
$V$ vs $V-I$ color-magnitude diagrams centered in the shells and supershells
of NGC~3109 compared with theoretical isochrones of Bertelli et al. (1994)
for $Z=0.001$.  The isochrones with ages $log~t= 6.6, 7.0, 7.8, 8.0, 8.7, 9.0, 9.7$
and $10.0$ years have been shifted according to  $(m-M)_0=25.62$ and
$E_{V-I}=0.05$ measured in this work.
}
\end{figure}

\begin{planotable}{lrllll}
\small
\footnotesize
\tablewidth{0pt}
\scriptsize
\tablecaption{Summary of parameters for NGC~3109}
\tablehead{
\multicolumn{1}{c}{Parameter}&
\multicolumn{1}{c}{Value}&
\multicolumn{1}{c}{}&
\multicolumn{1}{c}{Reference}}
\startdata
$(\alpha, ~\delta_{2000})$&$10:03:06.61, ~-26:09:32.0$& & 4 \nl
$(l, ~b)$                &$262.1^{\circ}, ~23.1^{\circ}$& & 4 \nl
Distance modulus         &$25.62\pm 0.05$ &    & 1,5 \nl
Distance                 &$1.33 ~Mpc$&          & 1,5 \nl
$Type$                   &  Sm IV &         & 4 \nl
$M_B$                    &  $-16.3$  &         & 2 \nl
$E(V-I)$                 &   0.05   &          & 1,5 \nl
b/a$_{disk}$             &   0.2    &          & 2 \nl
$V_{LG}$                 &   +129 km s$^{-1}$ & & 4 \nl
Inclination              &   $80^{\circ}$   &  & 2 \nl
Major axis                 &   17.4'   &      & 2 \nl
Minor axis                 &   3.5'    &      & 2 \nl
$Z_{disk}$               &   0.002  &          & 1,3 \nl
$Z_{halo}$               &   0.001  &          & 1,5 \nl
\enddata
\tablerefs{ \\
1. This work. \\
2. Jobin \& Carignan 1990. \\
3. Richer \& McCall 1995. \\
4. Sandage \& Carlson 1988. \\
5. Lee 1993. \\
}
\end{planotable}

\begin{planotable}{lrlllll}
\small
\footnotesize
\tablewidth{0pt}
\scriptsize
\tablecaption{Age Limits for HII Regions and Shells}
\tablehead{
\multicolumn{1}{c}{Object$^1$}&
\multicolumn{1}{c}{Xc,Yc}&
\multicolumn{1}{c}{Field}&
\multicolumn{1}{c}{$log ~Age ~(yr)$}&
\multicolumn{1}{c}{N stars}&
\multicolumn{1}{c}{Comments}}
\startdata
HII\# 1   & 1476, 1011  & 100$\times$100 pix& $>$7.5 & 83  &   \nl
HII\# 2   & 1330, 1061  & 100$\times$100 pix& $>$7.5 & 74  &   \nl
HII\# 3   & 1315,  832  & 100$\times$100 pix& $>$7.0 & 82  &   \nl
HII\# 4   & 1289, 1032  & 100$\times$100 pix& $>$7.5 & 72  &   \nl
HII\# 5   & 1227, 1087  & 100$\times$100 pix& $>$7.2 & 67  &   \nl
HII\# 6   & 1148,  938  & 100$\times$100 pix& $>$7.3 & 59  &   \nl
HII\# 7   & 1066, 1064  & 100$\times$100 pix& $>$7.2 & 91  & reddened?  \nl
HII\# 8   & 1031,  920  & 100$\times$100 pix& $>$7.2 & 67  &   \nl
HII\# 9   &  656, 1023  & 100$\times$100 pix& $>$7.7 & 77  & reddened?   \nl
HII\# 10  &  410,  750  & 100$\times$100 pix& $>$6.8 & 92  &   \nl
\nl
Shell\# 2    &  410,  748  & 100$\times$100 pix& $>$7.0 & 92  &   \nl
Shell\# 3    &  882,  755  & 100$\times$100 pix& $>$7.3 & 83  & supershell  \nl
Shell\# 4    & 1020, 1110  & 100$\times$100 pix& $>$7.7 & 90  & supershell  \nl
Shell\# 5    & 1020, 1110  & 100$\times$100 pix& $>$7.4 & 72  &   \nl
Shell\# 6    & 1290, 1218  & 100$\times$100 pix& $>$7.5 & 65  & mostly old stars \nl
Shell\# 7    & 1325, 1020  & 100$\times$100 pix& $>$7.2 & 78  &   \nl
Shell\# 8    & 1840, 1072  & 100$\times$100 pix& $>$7.5 & 60  &   \nl
Shell\# 9    & 1900,  800  & 100$\times$100 pix& $>$7.5 & 118  & center uncertain \nl
Shell\# 11   & 1788, 1266  & 100$\times$100 pix& $>$8.0 & 44  & only old stars  \nl
Shell\# 12   & 1321, 1325  & 100$\times$100 pix& $>$8.5 & 32  & only old stars  \nl
\enddata
\tablerefs{ \\
1. HII region IDs from Richer \& McCall (1992), \\
Shell IDs from Hunter et al. (1993).}
\end {planotable}

\end{document}